\documentclass[%
 reprint,
 superscriptaddress,
 longbibliography,
 amsmath,amssymb,
 aps,
 prapplied,
floatfix,
]{revtex4-2}

\usepackage{graphicx}% Include figure files
\usepackage{dcolumn}% Align table columns on decimal point
\usepackage{bm}% bold math
\usepackage[utf8]{inputenc}
\usepackage[T1]{fontenc}
\usepackage{mathptmx}
\usepackage{multirow}
\usepackage{hhline}
\usepackage{svg}
\usepackage[version=4]{mhchem}
\usepackage[english]{babel}
\usepackage{verbatim}
\usepackage{tabularx}
\usepackage{soul}
\usepackage[range-units = single]{siunitx}
\usepackage[hidelinks]{hyperref}
\usepackage{layouts}

\begin{document}

\title{Experimental characterization of photoemission from plasmonic nanogroove arrays}

\newcommand{\lbladdr}{LBNL, 1 Cyclotron Road, Berkeley, California 94720, USA}
\newcommand{\classeaddr}{CLASSE, Cornell University, 161 Synchrotron Drive, Ithaca, New York 14853-8001, USA}
\newcommand{\ncemaddr}{Department of Materials Science and Engineering, University of California, Berkeley, and National Center for Electron Microscopy, Molecular Foundry, Lawrence Berkeley National Laboratory, Berkeley, CA 94720}

\author{Christopher M. Pierce} \email{cmp285@cornell.edu} \affiliation{\lbladdr} \affiliation{\classeaddr}
\author{Daniel B. Durham} \affiliation{\ncemaddr}
\author{Fabrizio Riminucci} \affiliation{\lbladdr}
\author{Scott Dhuey}\affiliation{\lbladdr}
\author{Ivan Bazarov} \affiliation{\classeaddr}
\author{Jared Maxson} \affiliation{\classeaddr}
\author{Andrew M. Minor} \affiliation{\ncemaddr}
\author{Daniele Filippetto} \email{dfilippetto@lbl.gov} \affiliation{\lbladdr}

\date{\today}

%TC:ignore
\begin{abstract}
Metal photocathodes are an important source of high-brightness electron beams, ubiquitous in the operation of both large-scale accelerators and table-top microscopes. When the surface of a metal is nano-engineered with patterns on the order of the optical wavelength, it can lead to the excitation and confinement of surface plasmon polariton waves which drive nonlinear photoemission. In this work, we aim to evaluate gold plasmonic nanogrooves as a concept for producing bright electron beams for accelerators via nonlinear photoemission. We do this by first comparing their optical properties to numerical calculations from first principles to confirm our ability to fabricate these nanoscale structures. Their nonlinear photoemission yield is found by measuring emitted photocurrent as the intensity of their driving laser is varied. Finally, the mean transverse energy of this electron source is found using the solenoid scan technique. Our data demonstrate the ability of these cathodes to provide a tenfold enhancement in the efficiency of photoemission over flat metals driven with a linear process. We find that these cathodes are robust and capable of reaching sustained average currents over \SI{100}{\nano\ampere} at optical intensities larger than \SI{2}{\giga\watt\cm^{-2}} with no degradation of performance. The emittance of the generated beam is found to be highly asymmetric, a fact we can explain with calculations involving the also asymmetric roughness of the patterned surface. These results demonstrate the use of nano-engineered surfaces as enhanced photocathodes, providing a robust, air-stable source of high average current electron beams with great potential for industrial and scientific applications.
\end{abstract}
%TC:endignore
%TC:ignore
\maketitle
%TC:endignore

\section{Introduction}
High brightness electron sources for ultrafast applications require prompt emission of high-charge electron beams and direct injection into areas of extreme electromagnetic field amplitudes. 
Photoemission from metal surfaces has been the primary means of electron bunch generation, used by the large majority of user facilities around the world~\cite{lcls-ii_project_team_lcls-ii_2015,altarelli_xfel_2006, prat_thermal_2014}, owing to their fast response time and  robustness. %Indeed, the vacuum levels found in such extremely high field environments often do not favor the use of more delicate high quantum efficiency semiconductor materials.
Despite their broad use, metal cathodes have a few major disadvantages. First, the typical quantum efficiency for a metal exhibits values in the $10^{-5}$ region which, for high charge pulse extraction, requires laser pulse intensities close to the damage threshold of the material. With time and continuous operation, this has been shown to lead to partial ablation, increased surface roughness, and reduced brightness~\cite{akre_commissioning_2008}.
High intensities may also cause multi-photon absorption and photoemission, leading to the generation of unwanted halos, and an overall increase of beam thermal emittance~\cite{bae_brightness_2018}.
Furthermore, a typical metal work function requires UV photons for linear photoemission. 
The two-stage UV conversion from the initial infrared laser pulses has a substantial impact on the size and complexity of the photocathode laser system.
It may also impact the quality of the final pulse, resulting in substantial loss of energy,  degradation of transverse pulse shape, and limited control over longitudinal profile. 
Altogether, the low quantum efficiency and the high work function effectively limit the maximum average current that can be extracted by metal cathodes and, therefore, the range of applications of the relevant instrumentation. 

High quantum efficiency semiconductor films provide a possible path towards higher performance photocathodes. Depending on the choice of the material, the quantum efficiency can be orders of magnitude larger for a work function in the visible or infrared region~\cite{bazarov_thermal_2008}. 
Unfortunately, such cathodes are chemically reactive, and the vacuum levels found in high field photoinjectors often greatly complicate their use as high brightness electron sources.
Further, dark current may become an issue in those same systems for materials with an extremely low work function.

Nonlinear photoemission may offer another potential solution to avoid nonlinear wavelength conversion.
Depending on the material and laser parameters, it becomes more efficient to extract electrons from the cathode directly via multi-photon photoemission using infrared light, rather than perform wavelength conversion to  the UV~\cite{musumeci_multiphoton_2010}.
However, as is the case for linear photoemission, the small nonlinear yield of most flat metallic surfaces demands laser fluence values close to the material's damage threshold (typically on the order of \SIrange{0.1}{1}{\J \cm^{-2}}~\cite{kruger_femtosecond_2007}).

One path forward in improving the nonlinear yield of metals is by fabricating plasmonic structures by surface nanopatterning.
Nanoscale grooves formed on a gold photocathode have been shown to increase its nonlinear yield at \SI{800}{\nm} by up to six orders of magnitude~\cite{polyakov_plasmon-enhanced_2013}.
A similar concept using a grid of nanoscale holes showed a dramatic increase in the nonlinear yield of gold and copper photocathodes~\cite{li_surface-plasmon_2013,gong_high-brightness_2014}. On the other hand many questions remain open before such cathodes could be effectively considered as a reliable source for ultrafast application: Can we produce nano-engineered cathodes with repeatable properties? How does the mean transverse energy of the extracted beam depend on the nanostructures?  Can such structures provide stable high average currents for extended periods with no degradation?

In this work we provide a detailed characterization of nanogroove array photocathodes that demonstrates understanding of both the engineering and the physical aspects of this advanced class of electron photoemitters. 
First, in Sec.~\ref{sec:principles} we discuss the theory of plasmonic nanogroove photocathodes.
In Sec.~\ref{sec:fabrication} we explain the fabrication process, and confirm the design dimensions by direct measurements of their optical properties.
Nonlinear photoemission measurements performed on a \SI{20}{\kV} electron gun are reported in Sec.~\ref{sec:nonlinear-photoemission}. We find the non-linear photoemission coefficient for the nanostructured surfaces and are able to correlate its spread in values with the groove dimensions. We then confirm the polarization dependence of the photoemission, and perform continuous measurement of  average currents in excess of \SI{100}{\nano\ampere} to verify the enhanced electron yield and the photocathode stability. 
Lastly, in Sec.~\ref{sec:MTE} the mean transverse energy of the photocathode is characterized for different energies and the values found compared with the cathode's behaviour at the surface. 
The article then concludes by discussing future prospects for nanopatterned photoemitters.

\section{Principles of Plasmonic Nanogroove Photocathodes}
\label{sec:principles}
The ideal nanogroove cathode consists of a periodic array of trenches with depth ($d$) that extend infinitely in one direction and have nanometric width ($w$) in the other direction.
We define a coordinate system used for the rest of this paper with $\hat{z}$ pointing normal to the cathode surface, $\hat{y}$ running along the grooves, and $\hat{x}$ against the grooves.
Focusing for the moment on a single groove and imagining very large depth, light incident on the grooves may be coupled into modes within the gap that are best described by surface plasmon polaritons (SPP) within a metal-insulator-metal waveguide~\cite{le_perchec_why_2008} (the vacuum is the insulator in this case).
These SPPs require additional momentum to couple with free space illumination, owing to their dispersion relationship lying at larger wave-vector for the same energy than the light line.
For the case of the nanogrooves, the sharp edges at the entrance to the trenches can effectively provide such coupling~\cite{polyakov_plasmonic_2011}.
The corner's profile contains high spatial frequency components that allow light to diffract around it and onto the plasmon dispersion curve.

The finite depth of the groove acts to form a resonant Fabry-Perot-like cavity with the allowable modes determined by the depth, $d$.
The cavity depth that meets the resonance condition may be surprisingly small, only tens of \si{\nm} for infrared light.
This is explained by the fact that for the same energy, plasmons traveling along the walls of the gap may have an order of magnitude smaller wavelength than light in a vacuum~\cite{le_perchec_why_2008}.
The localization of optical energy to a nanometric region has the effect of field enhancement near the gap, which can exceed factors of one hundred and favor nonlinear photoemission.

\begin{figure}[t]
    \centering
    \includegraphics[width=\linewidth]{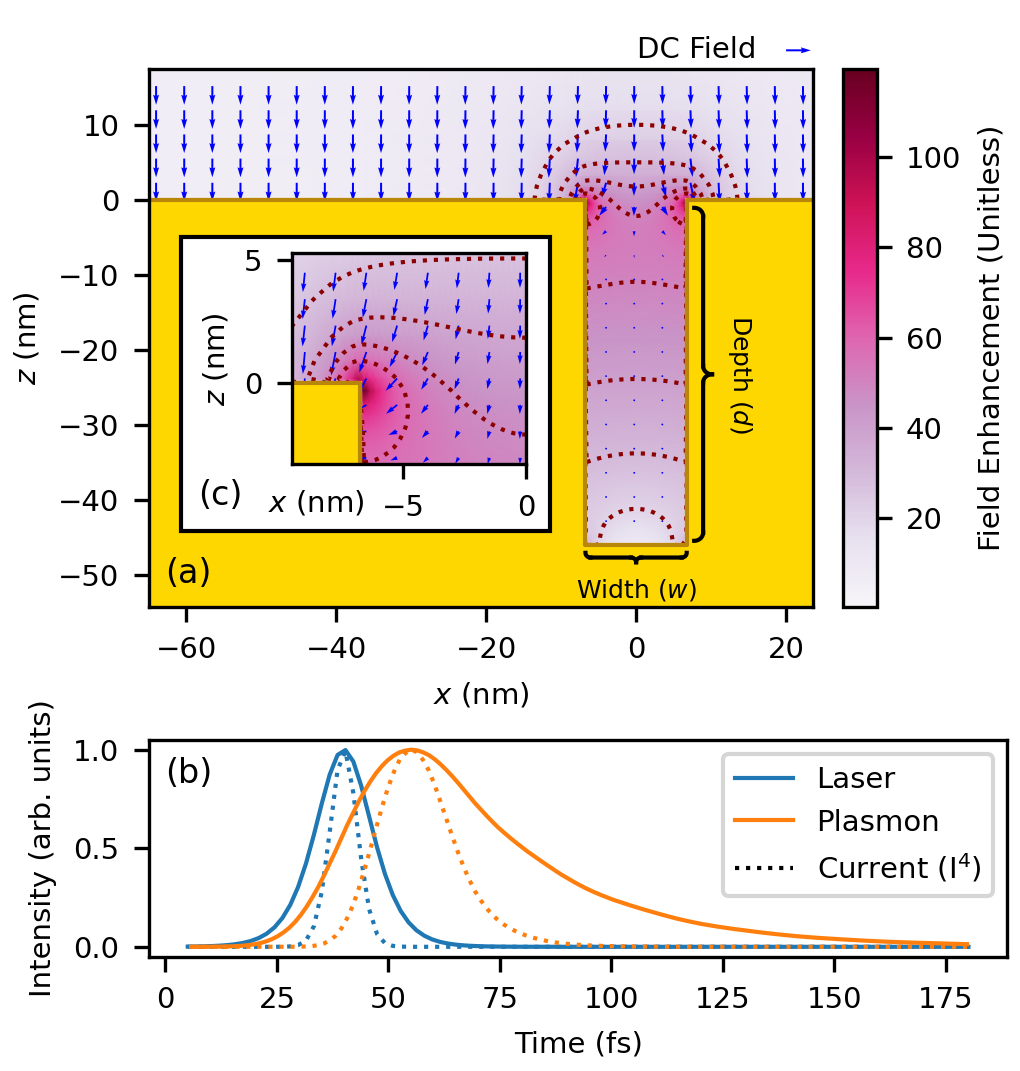}
    \caption{(a) Plot (with contours) of the field enhancement around a cross section of the nanogroove structure. Solution was computed with an FDTD code~\cite{noauthor_lumerical_2011}. The DC accelerating field, computed independently using a finite difference code~\cite{menzel_users_1987}, is shown as the blue arrows. Inset (labeled c) shows a magnified view of the groove edge; (b) Time response of structure to a \SI{15}{\fs} excitation computed with FDTD (plasmon) compared with flat surface (laser). Estimated current profiles are shown as dotted lines.}
    \label{fig:enhancement}
\end{figure}

Fig.~\ref{fig:enhancement}a shows an example of local optical field enhancement by a nanogroove cathode computed using a finite difference time domain (FDTD) code.
Specifically, Lumerical~\cite{noauthor_lumerical_2011} is used in this work.
Simulations are performed with periodic boundary conditions in the $\hat{x}$ direction and excited by linearly polarized plane waves.
The simulated cathode had grooves \SI{14}{\nm} wide, with a pitch of \SI{680}{\nm}, and was excited by light with a wavelength of \SI{770}{\nm}; representative of the cathodes studied in this paper.
The same picture also shows the computed local variation of a static externally applied electric field. 

The fact that emission occurs only at the sharp edges of the grooves may have an impact on the emittance of generated electron beams~\cite{li_surface-plasmon_2013} and high local optical intensity can damage the gold surface\cite{polyakov_plasmon_2012}.
However, the specific pattern used, the type and materials used during nano-fabrication, such as the sharpness of the pattern have an enormous impact on all of the above aspects. 

The high quality factor (and narrow bandwidth) of the plasmonic nanogroove also has consequences on the photocathode response time.
When the resonance bandwidth of the grooves is narrower than the bandwidth of the driving ultrafast laser, the field will continue to oscillate in the nanocavity longer than the duration of the excitation, effectively broadening the temporal response time of the cathode.
An example of this effect was computed for the nanogroove array photocathode in Fig~\ref{fig:enhancement}a by calculating the time-dependent field in response to excitation by ultrafast laser using an FDTD code (Lumerical~\cite{noauthor_lumerical_2011}).
The laser was \SI{15}{\fs} full-width-at-half-max (FWHM) and the calculated response of the structure was about \SI{42}{\fs} or a factor of three longer (Fig.~\ref{fig:enhancement}b).
The approximate bandwidth ($\Delta \lambda$) and peak absorption wavelength ($\lambda$) were \SI{15}{\nm} and \SI{770}{\nm} FWHM.
The minimum time-bandwidth product of an approximately Gaussian pulse is 0.44~\cite{diels_ultrashort_2006}.
From this, an estimate of the optical time response can be calculated as
\begin{equation}
    \Delta t_{\text{Response}} =\sqrt{\Delta t_{\text{Laser}}^2 + (a_1a_2 \lambda^2 / (4\pi c \Delta\lambda))^2},
    \label{eq:response}
\end{equation}
where $a_1$ and $a_2$ are conversions from RMS to FWHM values of the absorption bandwidth and nanogroove time response respectively. Assuming Gaussian profiles, both $a_1$ and $a_2$ in Eq. \ref{eq:response} are approximately equal to 2.35, leading to a final cathode temporal response of \SI{\sim50}{\fs} FWHM.
In order to obtain the extracted electron beam pulse duration, one would have to take into consideration the particular photoemission order used. In our case, the current density is proportional to the fourth-order of laser intensity, which suppresses the tails of the optical response and shrinks the final duration by a factor of two (for a Gaussian-like pulse).
For ultrafast applications, other plasmonic cathode schemes that do not rely on resonant cavities may support higher bandwidths and allow use with even shorter laser pulses~\cite{durham_plasmonic_2019}.

\section{Cathode Fabrication and Optical properties}
\label{sec:fabrication}
Photocathodes were fabricated out of gold using the template stripping method~\cite{vogel_as_2012,polyakov_light_2011}.
While plasmonic structures can be fabricated using other methods such as focused ion beam milling and the lift-off procedure, template stripping has been shown to yield superior surface roughness.
Prior work has found RMS roughness of \SI{0.2}{\nm} RMS for template stripping compared to \SI{1.4}{\nm} for thermally evaporated metals~\cite{vogel_as_2012}).
A silicon wafer was UV/ozone cleaned for 5 minutes and spin coated with HSQ 2\% resist.
It was baked at \SI{100}{\degreeCelsius} for \SI{1}{\minute} and then patterned with electron beam lithography.

\begin{figure}[t]
    \centering
    \includegraphics[width=\linewidth]{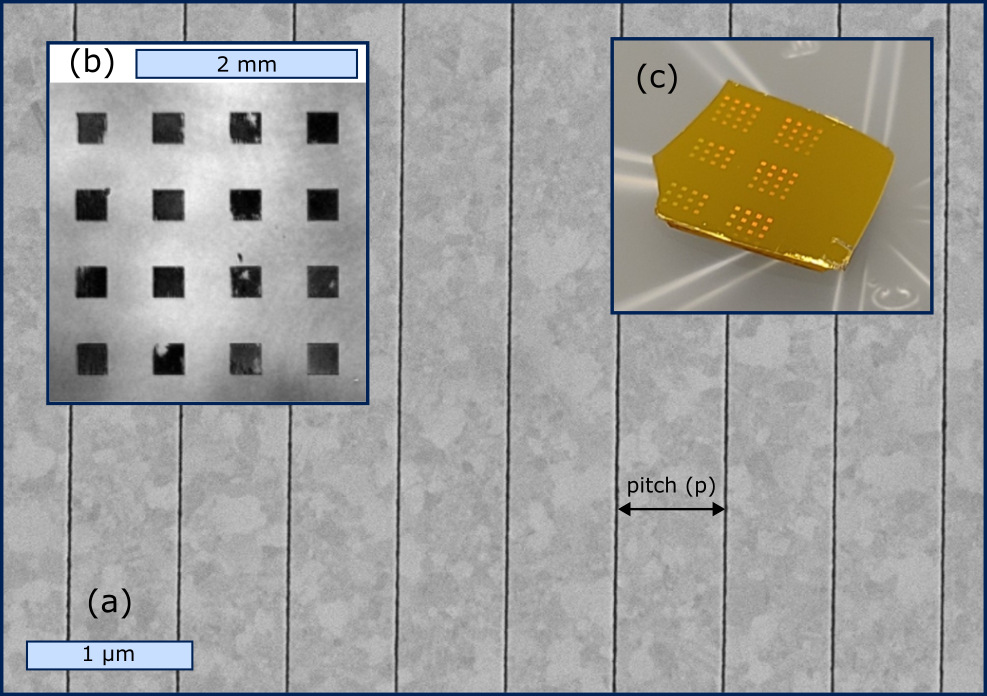}
    \caption{(a) Scanning electron micrograph of the nanopatterned cathode; (b) spatial image of reflectance of 16 plasmonic nanogroove photocathodes contained in a single square of the sample; (c) An image of the fabricated cathode.}
    \label{fig:micrographs}
\end{figure}

Cathodes with varying geometries are arranged in a square grids (a 4x4 pattern), and multiple grids are imprinted along a single wafer, with an edge-to-edge distance of about \SI{1}{\mm} (see Fig.~\ref{fig:micrographs}b,c).
The geometric dimensions of each of the 16 cathodes within a single square are varied, with a different groove pitch for each row, and a different width for each of the four columns.
The groove width was varied in part by controlling the electron dose, leaving some calibration required for this dimension.
The groove depth was fixed by the fabrication procedure at \SI{50}{\nm}.
After exposing the resist, the template was cleaned using RIE oxygen plasma for \SI{30}{\second} and \SI{150}{\nm} of gold was deposited.
UV curable epoxy was used to adhere a thin glass substrate to the gold and pressure caused this assembly (substrate, epoxy, and gold) to separate from the template revealing the nanopatterned cathode.

\begin{figure}[t]
    \centering
    \includegraphics[width=\linewidth]{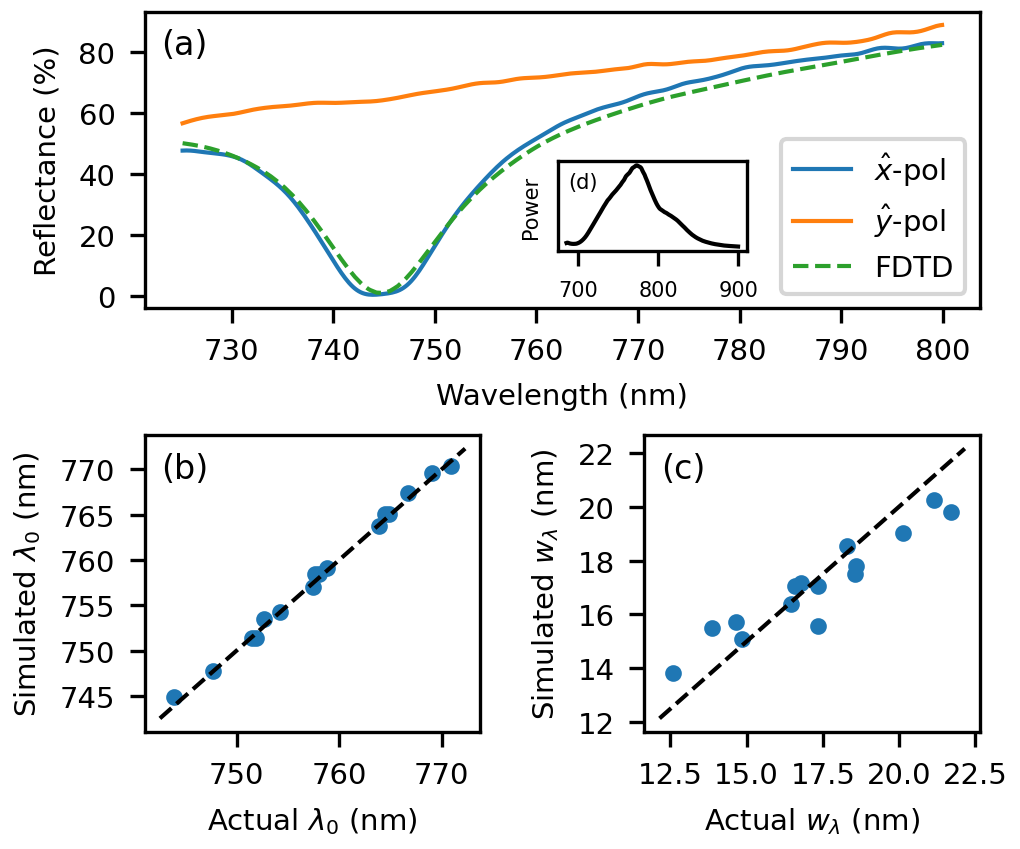}
    \caption{(a) An example of reflectance spectra measured from one of the cathodes (\SI{670}{\nm} pitch, \SI{17.1}{\nm} width) with a fit to the FDTD model using groove width and cathode tilt as the free parameter; (b) Peak absorption wavelength of model and measured grooves; (c) Full width at half max of absorption peak in model and fabricated sample. (d) Spectrum of the mode-locked driving laser.}
    \label{fig:reflectance-spectra}
\end{figure}

The quality of the fabrication was first verified by imaging the surface via scanning electron microscopy (Fig.~\ref{fig:micrographs}a), and confirming the close match between the array dimensions and the target values.
The most central square grid was then used for optical and photoemission measurement as it was the easiest to align along the axis of the photoemission setup.  

We then performed reflectivity measurements, starting with near-IR imaging, of all the 16 patterns across the selected square in the sample. 
We used a \SI{770}{\nm} centered non-modelocked Ti:Sapphire laser oscillator as illumination source.
The linear laser polarization was tuned to point in the direction across the grooves (IE $\hat{x}$), while the laser pulse hit the cathode at normal incidence.
As can be qualitatively seen in Fig.~\ref{fig:micrographs}b, we observed strong suppression of the reflectivity in the regions that contain the nanopatterning.

The laser system was then mode-locked and its full bandwidth was used to measure the reflectance spectra of the nanogroove photocathodes (see Fig.~\ref{fig:reflectance-spectra}d).
Difference spectra were calculated using the beam reflected from the patterned surface against a reference pulse from an upstream 50-50 beamsplitter for directions of linear polarization pointed along $\hat{x}$ and $\hat{y}$.
The reflected spectra for $\hat{x}$ polarized light (IE against the grooves) were fit to FDTD calculations~\cite{noauthor_lumerical_2011} using the groove width for each of the 16 samples and a single overall angle of incidence as the free parameters.
We found the best-fit angle of incidence for the cathode to be \SI{0.5}{\degree}.
Fig.~\ref{fig:reflectance-spectra}a reports an example of fit result for one cathode, exemplifying the close match found between the simulated and measured reflectance. The groove widths extracted are reported in Tab.~\ref{tab:table1} ( the fourth column) for all the cathodes in a square. From such results we are able to confirm our fabrication methodology, as width values increase with column indices, i.e. with electron beam lithography dose, matching our expectations.
Measured and simulated fit peak absorption wavelength and full width at half max of the peak are compared in Fig. ~\ref{fig:reflectance-spectra}b and c, showing excellent agreement.

Overall, these measurements demonstrate an ability to fabricate nanopatterned photocathodes with engineered optical properties. 

\begin{table}[b]
\caption{\label{tab:table1}
Dimensions of the nanogroove cathodes: pitch ($p$) and groove width extracted from the fit of the reflectance spectra ($w$). Measured effective non-linear yield ($a_4$) and lower bound placed on enhancement of nonlinear yield over flat gold.
}
\begin{ruledtabular}
\begin{tabular}{cccccc}
\textrm{Row}&
\textrm{Col}&
\textrm{$p$ (\si{\nm})}&
\textrm{$w$ (\si{\nm})}&
\textrm{$a_4$ $\left(\si{(\cm^2 \A^{-1})^4}\right)$}&
\textrm{Enhancement Bound}\\
\colrule
1 & 1 & 670 & 14.5 & $3.0 \times 10^{-37}$ & $7.5 \times 10^{5}$\\
1 & 2 & 670 & 14.5 & $4.0 \times 10^{-38}$ & $1.0 \times 10^{5}$\\
1 & 3 & 670 & 15.8 & $2.3 \times 10^{-39}$ & $5.7 \times 10^{3}$\\
1 & 4 & 670 & 17.1 & $1.6 \times 10^{-38}$ & $4 \times 10^{4}$\\
2 & 1 & 680 & 14.5 & $8.9 \times 10^{-37}$ & $2.2 \times 10^{6}$\\
2 & 2 & 680 & 14.5 & $1.3 \times 10^{-37}$ & $3.3 \times 10^{5}$\\
2 & 3 & 680 & 15.4 & $1.0 \times 10^{-36}$ & $2.5 \times 10^{6}$\\
2 & 4 & 680 & 16.2 & $2.9 \times 10^{-37}$ & $7.3 \times 10^{5}$\\
3 & 1 & 690 & 14.1 & $3.0 \times 10^{-37}$ & $7.5 \times 10^{5}$\\
3 & 2 & 690 & 14.1 & $2.2 \times 10^{-37}$ & $5.5 \times 10^{5}$\\
3 & 3 & 690 & 15.4 & $2.2 \times 10^{-37}$ & $5.5 \times 10^{5}$\\
3 & 4 & 690 & 15.8 & $6.4 \times 10^{-39}$ & $1.6 \times 10^{4}$\\
4 & 1 & 700 & 14.1 & $4.5 \times 10^{-38}$ & $1.1 \times 10^{5}$\\
4 & 2 & 700 & 13.7 & $1.8 \times 10^{-37}$ & $4.5 \times 10^{5}$\\
4 & 3 & 700 & 14.5 & $5.8 \times 10^{-40}$ & $1.5 \times 10^{3}$\\
4 & 4 & 700 & 15.4 & $3.7 \times 10^{-39}$ & $9.2 \times 10^{3}$\\
\end{tabular}
\end{ruledtabular}
\end{table}

\section{Nonlinear Photoemission from Nanopatterned Cathode}
\label{sec:nonlinear-photoemission}
In this section we describe the measured non-linear electron yield and average current of the nanogroove arrays. 

The generalized Fowler-DuBridge model of multi-photon photoemission gives the scaling of n-photon current density, $J_n$, with laser intensity as the following~\cite{musumeci_multiphoton_2010, ferrini_non-linear_2009}.
\begin{equation}
    J_n = a_n A_0 \left(\frac{e}{h\nu}(1-R_\nu)I\right)^nT^2F\left(\frac{nh\nu - e\phi}{k_BT}\right),
    \label{eq:fowler-dubridge}
\end{equation}
where $a_n$ is a cathode dependent constant representing the chance of multiphoton excitation, $h$ is Planck's constant, $k_B$ is the Boltzmann constant, $e$ is the fundamental charge, $\nu$ is the optical frequency, $R_\nu$ is the metal's reflectivity, $\phi$ is the work function, $I$ is the optical intensity, $T$ is temperature, and $n$ is the order of emission.
The value $A_0 = 4\pi m_e k_B^2 e/h^3\approx\SI{120}{A \cm^{-2}\K^{-2}}$ is the Richardson constant with $m_e$ as the electron mass.
The Fowler function can be written as $F(x) = \int_0^\infty\mathrm{d}y~\ln(1 + \exp{(-y-x)})$.

The literature value of $\phi$ for gold is \SI{5.4}{\eV}~\cite{bass_handbook_2000}. Therefore we expect fourth order photoemission from the cathode when using \SI{800}{\nm} (\SI{1.54}{\eV}) photons.
Typical values for $a_4$ of flat gold~\cite{schweikhard_multiphoton_2011} lie around $a_4 \approx \SI{e-43}{(\cm^2 \A^{-1})^4}$.
Previous work on nanopatterned gold has demonstrated non-linear electron yield enhancements of the order of $10^6$ with pA-scale currents~\cite{polyakov_light_2011}.

\begin{figure}[t]
    \centering
    \includegraphics[width=\linewidth]{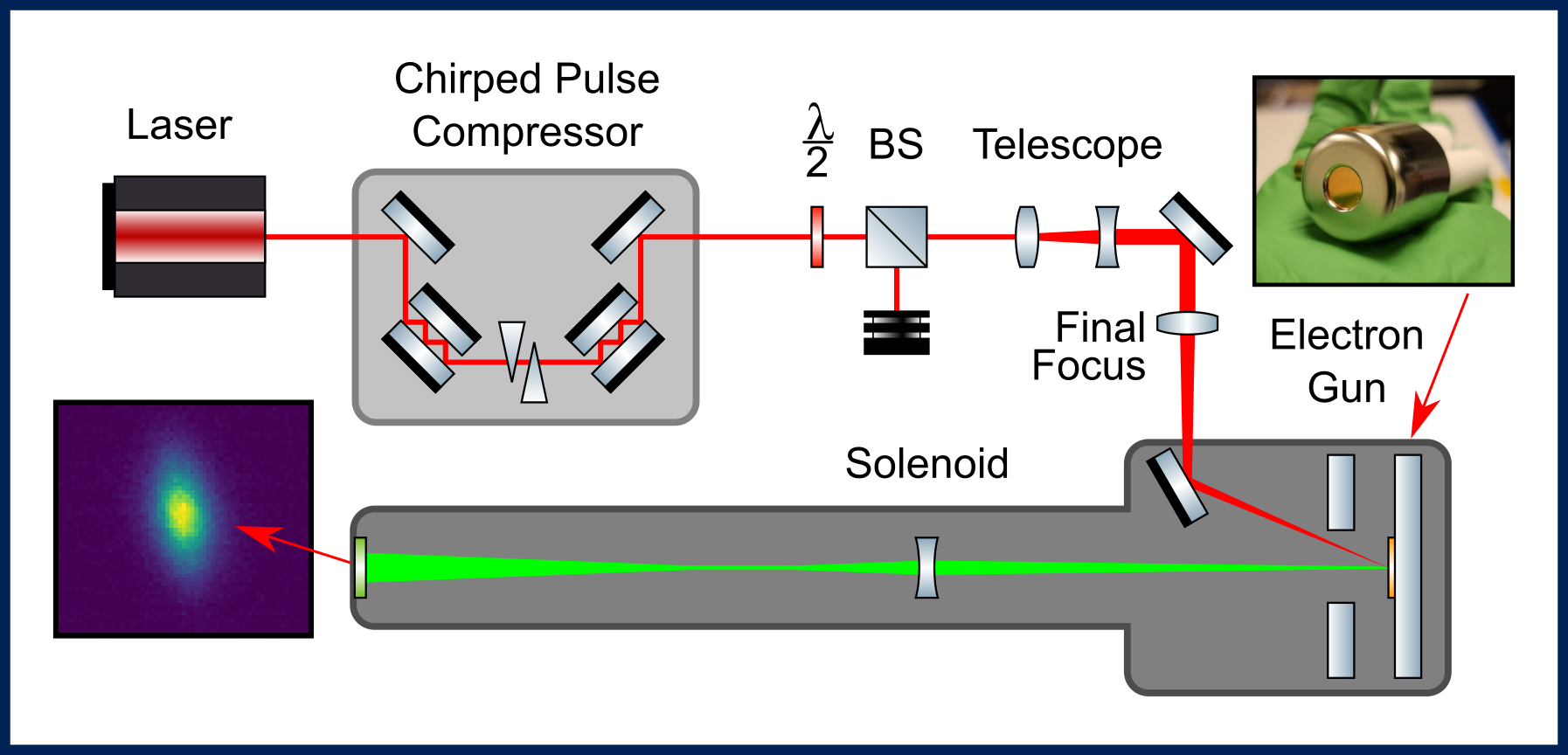}

    \caption{
    A schematic of the beamline used in the measurements.
    An \SI{80}{\MHz} Ti:Sapphire oscillator emits pulses centered around \SI{770}{\nm}. The light is sent through a chirped pulse compressor for temporal compression, and the beam is focused to a waist using a \SI{600}{\milli\meter} focal length lens placed just before the window of the vacuum chamber.
    Intensity may be adjusted with a beamsplitter (BS) and half-waveplate ($\lambda/2$) pair.
    A telescope is used to expand the beam before hitting the lens to decrease its ultimate focused size.
    The electron beam emitted from the cathode is accelerated and sent through a solenoid lens before being imaged on a scintillator screen.
    }
    \label{fig:beamline}
\end{figure}

A schematic of our experimental setup for the measurement of electron beams from nanostructures is shown in Fig.~\ref{fig:beamline}.
We transferred the nanopatterned wafers into a \SI{20}{\kV} electron gun, and used the \SI{80}{\MHz} repetition rate mode-locked Ti:Sapphire oscillator as the drive laser. 
The pulse was sent through a chirped pulse compressor to achieve a Fourier-transform-limited pulse length of \SI{\sim15}{\fs} at the sample, also confirmed by autocorrelation measurements. 
The pulse was then focused to an RMS spot size \SI{40}{\um} at the cathode plane with a small angle of incidence of 4 degrees with respect to the surface normal.
The intensity was varied using the combination of an achromatic half-wave plate and a polarizing beamsplitter. 
The maximum laser energy that could be sent to the cathode after transport, and longitudinal and transverse shaping was \SI{1}{\nano\joule}.

\begin{figure}[t]
    \centering
    \includegraphics[width=\linewidth]{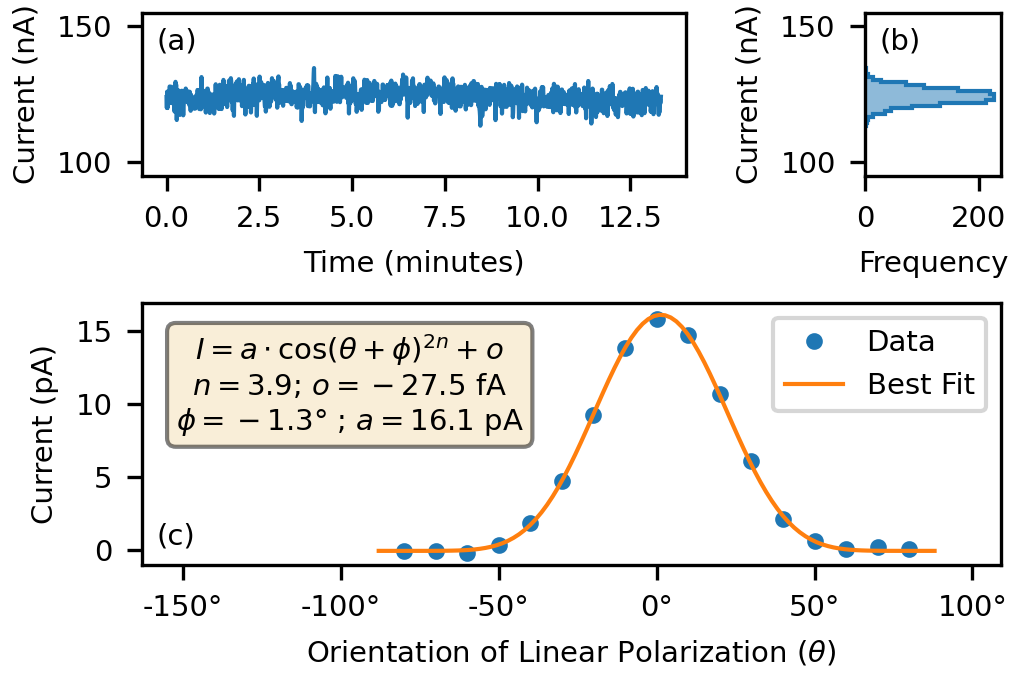}

    \caption{(a) Time series measurements of photocurrent from a nanopatterned photocathode; (b) Histogram of jitter in photocurrent; (c) Measurements of photocurrent as angle of linear polarization is changed with line of best fit.}
    \label{fig:stability}
\end{figure}

To begin, the polarizing beamsplitter was temporarily removed and the half-wave plate was used to control the orientation of linear polarization of the laser.
Photocurrent was measured from a single cathode using a lock-in amplifier and fit to the model $J(\theta) = A\cdot(\cos^2(\theta + \phi))^n + o$ where $A$ is the amplitude, $o$ is an offset, and $\phi$ is a phase to account for mis-positioning of the half-wave plate in its rotation mount.
The angle, $\theta$, is oriented so that \SI{0}{\degree} is close to $\hat{x}$ in the coordinate system of the cathode.
Our data and line of best fit are shown in Fig.~\ref{fig:stability}c and we conclude from the goodness of fit that only the polarization of light running "against the grain" of the grooves is able to excite plasmons and cause multiphoton photoemission, as expected.

\begin{figure*}[t]
    \centering
    \includegraphics[width=\linewidth]{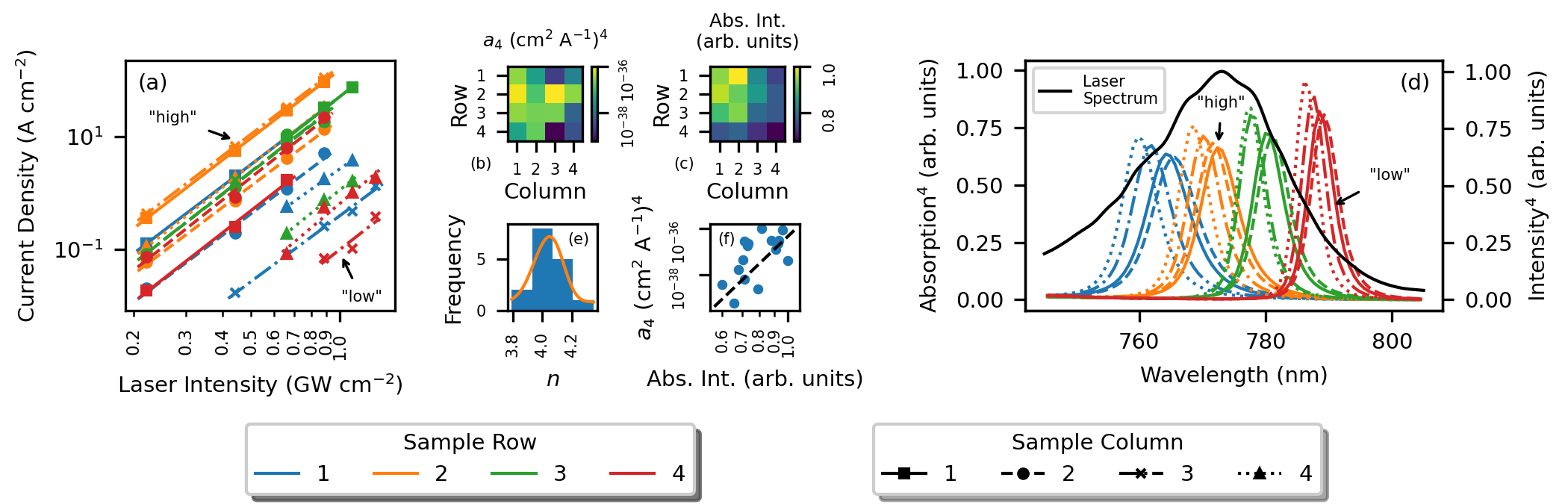}
    \caption{
    (a) Measurements of photocurrent as laser intensity is varied with power-law fits shown as lines; (b) Best fit yield ($a_n$ in Eq. \ref{eq:fowler-dubridge}), laid out by location on the 4x4 grid of cathodes (Fig.~\ref{fig:micrographs}b); (c) estimate of absorbed laser intensity, laid out by location in grid, log-scaled; (d) fourth power of nanogroove absorption overlaid with laser's spectrum; (e) The distribution of best-fit exponents in power-law curves (blue histogram, orange kernel density estimator); (f) Scatter plot of yield ($a_4$) and estimated absorbed laser intensity; colors and line style in (a) and (d) indicate sample location and high/low yield examples are highlighted.).
    }
    \label{fig:nonlinear-yield}
\end{figure*}

The polarizing beamsplitter was replaced and emitted electron photocurrent was measured as a function of optical intensity (Fig.~\ref{fig:nonlinear-yield}a) for each of the cathodes in a square within the wafer. Eq.~\ref{eq:fowler-dubridge} was then used in a fit to find the nonlinear yield coefficient ($a_4$), using 98\%~\cite{bass_handbook_2000} as the value of gold's reflectivity at \SI{760}{\nm}.

The distribution of measured non-linear yield exponents is shown in Fig.~\ref{fig:nonlinear-yield}e. All of the measured behaviours are consistent with fourth order photoemission.
The measured value of nonlinear yield coefficient varied from $a_4$ = \SIrange{3e-40}{6e-37}{(\cm^2 \A^{-1})^4}. A representation of its value distribution across the 16 cathodes in a square is shown in Fig.~\ref{fig:nonlinear-yield}b.

One explanation for the variation in $a_4$ is the change in optical response of the grooves depending on their geometry (see Fig.~\ref{fig:reflectance-spectra}a), that will change the overall coupling of the laser's power into the structure.
Depending on the value of the peak absorption wavelength, the reflectance curve will align better or worse with the power spectrum  of the laser (Fig.~\ref{fig:nonlinear-yield}d),
changing the amount of total absorbed intensity from the laser that then excites electrons in the metal.
The yield should be related to the fourth power of the absorbed intensity which can be estimated as the integral,
\begin{equation*}
    (\text{Abs. Int.})^4 = \left[\int_0^\infty I(\lambda)\cdot A(\lambda)\mathrm{d}\lambda\right]^4,
\end{equation*}
where $I(\lambda)$ is the laser's spectrum and $A(\lambda)$ is the nanogroove's absorption spectrum.
Indeed, if we evaluate this integral numerically (Fig.~\ref{fig:nonlinear-yield}c) for each groove array, we observe a correlation (Fig.~\ref{fig:nonlinear-yield}f).

Current emission from the flat (non-patterned) gold surface was below our measurement sensitivity, owing to the limited available optical intensity in the setup.
Nevertheless, with a maximum power density achievable of \SI{\sim2}{\giga\watt \cm^{-2}}, and a  measurement system's noise floor of \SI{\sim50}{\femto\ampere}, we can calculate an upper bound for the nonlinear yield coefficient of $a_4 < \SI{4e-43}{(\cm^2 \A^{-1})^4}$, with a yield enhancement from nanopatterning in excess of $10^6$.
This bound agrees with previous measurements on identically prepared flat gold that found $a_4 = \SI{1e-43}{(\cm^2 \A^{-1})^4}$~\cite{durham_design_2020}.

The current stability of the cathode performance is summarized in Fig.~\ref{fig:stability}.
A continuous acquisition of average electron current values over about \SI{12}{\minute} was performed, with the laser pulses delivering the maximum available energy (Fig.~\ref{fig:stability}a). 
A stable average current value of \SI{120}{\nano\ampere} was measured, with fluctuations measured to be 2.3\% (see Fig.~\ref{fig:stability}b).

This experiment exemplifies the disruptive potential of the technology in high average current electron sources. Indeed, assuming
a gold UV quantum efficiency of $10^{-5}$~\cite{fischer_uv_1988} and a typical conversion efficiency from NIR to UV of 7.5\%, then linear photoemission would require a tenfold increase in laser power to generate the same average current, about \SI{0.75}{\watt} in the NIR against the \SI{80}{\milli\watt} used in the experiment. 

\section{Mean Transverse Energy of the Emitted Electron Beam}
\label{sec:MTE}
In this section we explore the mean transverse energy (MTE)~\cite{lee_review_2015} of the nanogroove cathodes.
After showing that surface nanopatterning  can lead to very large nonlinear yield enhancements in metals, we now seek to explore the effects of such enhancement on the beam's transverse brightness.
Early work on nanopatterned photocathodes found larger emittance values than what is expected from a flat surface~\cite{li_surface-plasmon_2013}. 
Since then, substantial work tailored at improving the photoemission properties of these cathodes  has been carried out, for example by selecting and optimizing the fabrication technique for minimal roughness and sharp patterns, and by developing methods for in-situ optical characterization of the structures~\cite{durham_plasmonic_2019}.
Owing to such developments, we are now able to relate photoemission properties such as the cathode electron beam MTE, to the surface mechanical and optical characteristics.

\begin{figure}[t]
    \centering
    \includegraphics[width=\linewidth]{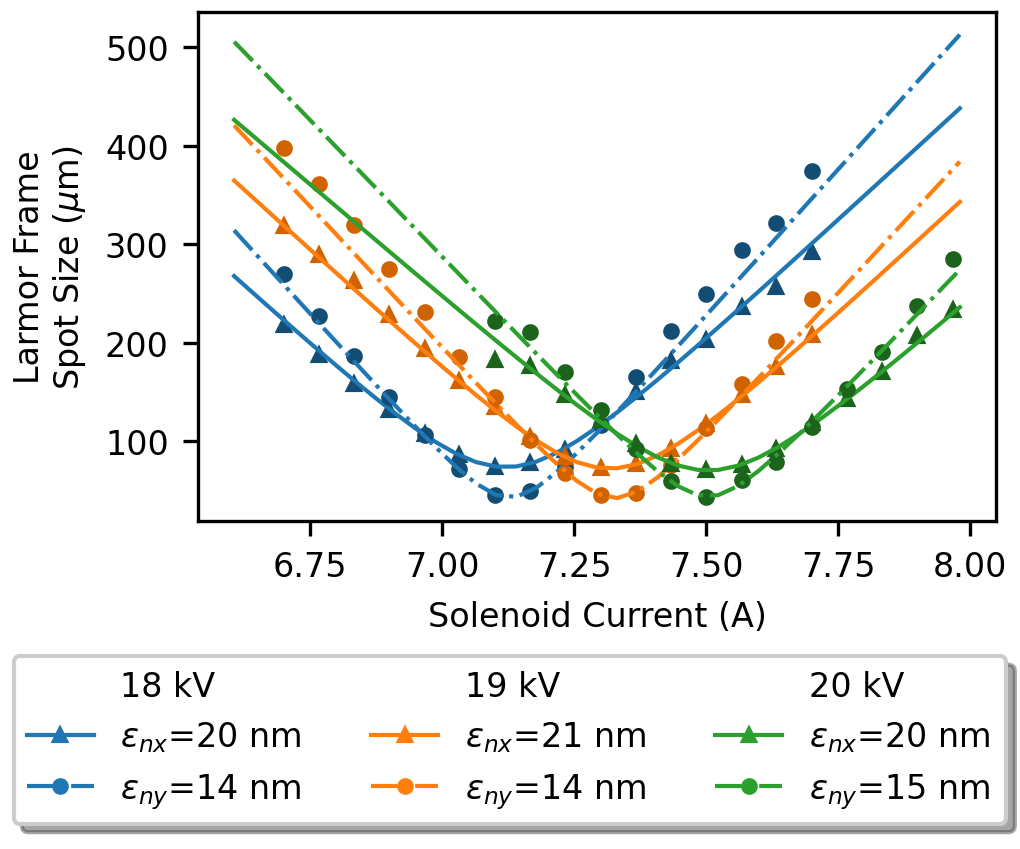}

    \caption{Solenoid scan measurements of the generated beam's normalized emittance. 
The beam's size (shown here in the Larmor frame) is measured as a function of current in the beamline's solenoid. 
Fits to a linear model of the beamline are shown as the curves with the best fit emittance in the legend.}
    \label{fig:solenoid-scans}
\end{figure}

We investigate the transverse emittance of the grooves using the solenoid scan method.
The electron gun is biased at high voltage to generate a beam from one of the cathodes in the square (row two and column one in Fig.~\ref{fig:micrographs}b; \SI{680}{\nm} pitch and \SI{14.5}{\nm} width).
The generated electron beam passes through a solenoid lens a few centimeters away from the cathode and hits a scintillator screen \SI{60}{\cm} away. Here the beam's RMS spot size was measured as function of strength of the solenoid lens.
The experiment was repeated at \SI{20}{\kV}, \SI{19}{\kV}, and \SI{18}{\kV} (data shown in Fig.~\ref{fig:solenoid-scans}) and the beam sizes were fit using a linear model of transport including the accelerating electric field in the gun, following the procedure in~\cite{bazarov_thermal_2008, gulliford_new_2012} to recover the initial phase space moments.
The laser RMS spot size at the cathode was measured to be \SI{20}{\um}. We find the cathode MTE to be asymmetric, with  MTE$_x$ = \SI{510}{\meV} and MTE$_y$ = \SI{250}{\meV}.
The x and y axes are as before with with $\hat{x}$ running against the grooves and $\hat{y}$ running along them.
The MTE along $\hat{y}$ is close to what's typical for a flat metal with this excess energy~\cite{dowell_quantum_2009}.
On the other hand, the MTE in the horizontal ($x$) plane shows a substantial increase, which we attribute to geometric effects, as described below.

While nanopatterning has clear benefits for the non-linear yield of photocathodes, those same nanoscale features are a form of surface roughness.
Surface roughness is well known to cause an increase in the MTE of the emitted electrons from a photocathode~\cite{bradley_transverse_1977}.
Two major effects contribute to this increase: the additional transverse momentum gained from the local distortions of electric fields around surface features (as in Fig.~\ref{fig:enhancement}) and the local deviation of the average direction of photoemission with respect to the global beamline axis, which follows the surface normal.
For nanogroove photocathodes, these effects will vanish in one direction ($\hat{y}$ in our setup), thanks to the structure's translational symmetry.
To understand the increase of emittance on the horizontal ($x$) axis, we will now estimate the contribution of both effects in our setup. 

\begin{figure}[t]
    \centering
    \includegraphics[width=\linewidth]{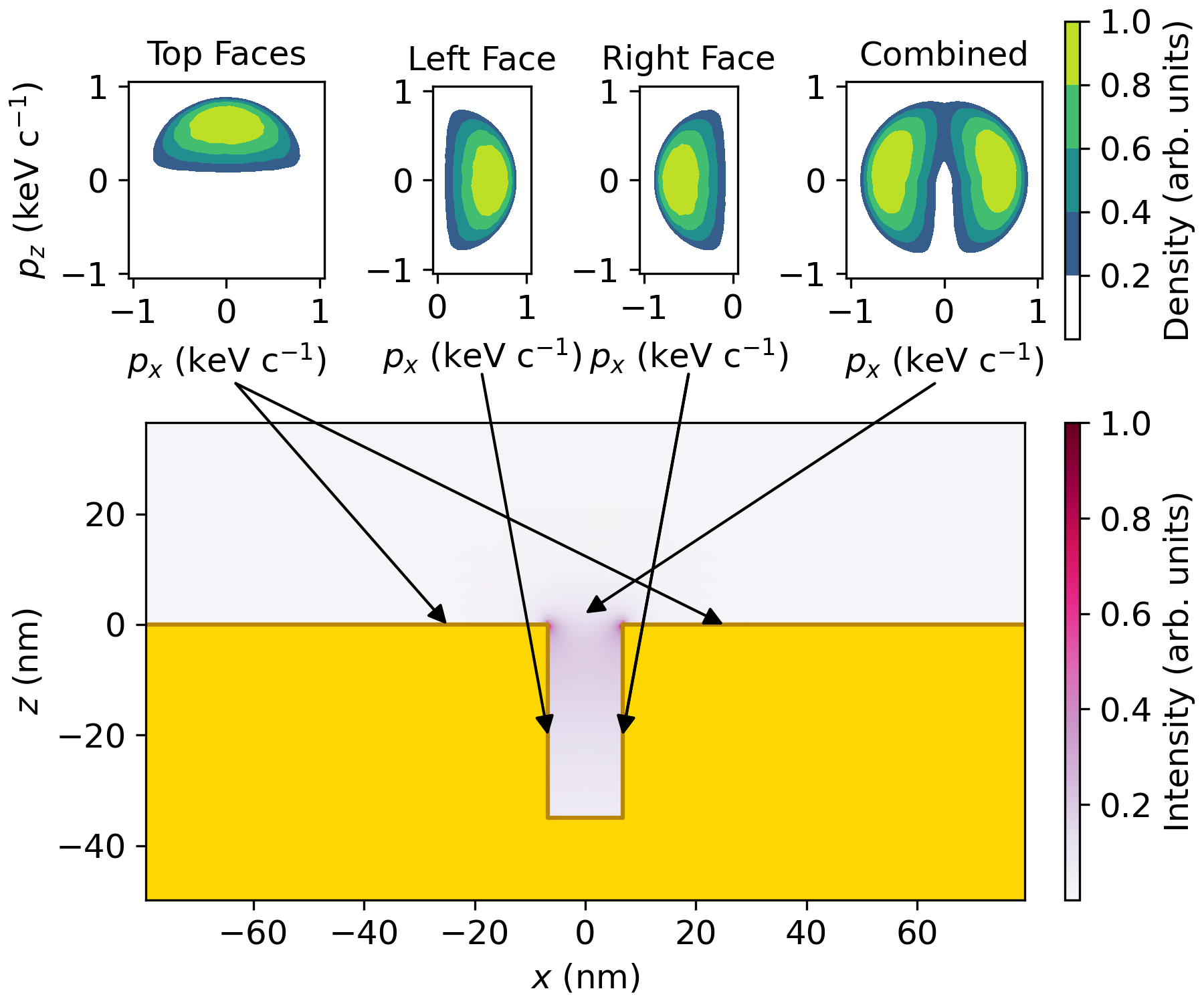}

    \caption{
    A schematic showing how the initial momentum distribution from each face of the nanogroove may be combined to form an estimate of MTE including roughness effects.}
    \label{fig:roughness}
\end{figure}

For the first effect, we compute the value of the externally applied electric field around the nanogrooves using the finite difference method~\cite{menzel_users_1987}, assuming perfect edges (i.e. a radius of curvature equal to the simulation mesh size).
This is shown as the blue arrows in Fig.~\ref{fig:enhancement}a.
We then compute the integral of the transverse field along the particle trajectory starting from the groove edge to find the maximum transverse energy acquired by the electrons. For the max achievable cathode field in the gun (which will maximize this effect) of \SI{7}{\mega\volt \meter^{-1}}, this effect only adds \SI{\sim10}{\meV} of transverse energy to the emitted electrons.

To estimate the effect of the local variations of the surface normal on the MTE, we start by randomly sampling the 3D momentum distribution of electrons emitted from a flat metallic surface~\cite{dowell_quantum_2009}, generating a set of 100k virtual particles.
Assuming a work function for gold of \SI{5.4}{\eV}~\cite{michaelson_work_1977} and fourth order photoemission process, a numerical calculation of the MTE of these particles gives a value of \SI{257}{\meV}, in close accordance with our measurement in the vertical plane and the analytical expression $\text{MTE} = (nh\nu - \phi)/3$.
Rotating the distribution following the local normal to the surface, and adding together all of the contributions, we can obtain an estimate of the total MTE including the effect of nanopatterning.
The number of electrons emitted from each nanogroove face along the surface is weighted by the integral of the fourth power of the intensity along it, extracted by FDTD simulations (Fig.~\ref{fig:enhancement}a).
This gives a ratio of side wall to top emission of 6.9:1 and the total estimated MTE with normal vector effects included of \SI{481}{\meV}.
Adding in the \SI{\sim 10}{\meV} calculated above due to the effects of the transverse fields, this approaches our measured value in the horizontal plane of \SI{510}{\meV}.
A visual of how this estimate is made can be found in Fig.~\ref{fig:roughness}.

\section{Conclusion}

In this work we report the development and engineering of a nanopatterned metal photocathode for high brightness ultrafast electron generation. 
We demonstrate increased average current when using plasmon-assisted multiphoton photoemission with respect to linear photoemission in our setup, overcoming the major drawbacks of metal cathodes caused by their poor QE in the UV, and paving the way to their use in high average current-high brightness applications, such as X-FELs and UED setups.

Our fabricated cathodes closely match their designed optical performance. 
In particular, we are able to tune the peak absorption wavelength of the structure to the spectral peak of the driving laser.

Electron yield was strongly enhanced via surface nano-structuring, a factor in excess of $10^6$ over fourth-order photoemission from flat gold, and a reduction in power by a factor of \SI{\sim 10}{} compared to linear photoemission for the intensities achieved in this work.  Continuous operations at high average current showed no degradation (Fig.~\ref{fig:stability}a).
To showcase the potential impact of such nanopatterned cathodes, we compare their requirements with typical metal photocathodes used in large-scale facilities. As an example, The LCLS X-FEL at SLAC \cite{emma_first_2010} uses linear photoemission from flat copper cathodes.
By using the operational values for the laser, cathode QE, and beam charge \cite{zhou_impact_2012,zhou_establishing_2015} (\SI{3}{\ps} FWHM, \SI{150}{\pico\coulomb} pulse from a \SI{1}{\mm} hard edge spot size, QE of \SI{4e-5}{}) and a typical operational conversion efficiency from IR to UV of 7.5\%, a flat copper cathode requires about \SI{240}{\micro\joule} of energy in the IR pulse, compared to the \SI{8}{\micro\joule} necessary for the gold nanostructured photocathode presented here. The advantage of nanopatterning becomes even more pronounced in applications requiring low charge and femtosecond-long pulses, such as UED setups \cite{filippetto_ultrafast_2022-1}. The example shown in Fig.~\ref{fig:stability}a, \SI{1.5}{\femto\coulomb} electron beams are produced using only \SI{1}{\nano\joule} of IR energy extracted directly from an ultrafast laser oscillator. 

The normalized transverse emittance of the photoemitted beam was measured systematically for different beam energies, providing a benchmark value for the transverse brightness of nanopatterned cathodes. 
The measured asymmetry in the emittance can be fully explained by the geometry of the structure, with its asymmetric roughness.
These results suggests an interesting application of nanogroove arrays as a future platform for studying the effects of roughness on electron source brightness
In these systems, the roughness can be engineered to take on certain profiles.
Further, the fact that the cathode "acts flat" in one direction provides a control measurement to directly compare the effects of roughness against in each sample.
For the application of cathodes in high brightness photoinjectors, although the emittance is increased in the direction normal to the grooves, it is still in line or better than typical values measured in ultrafast X-ray user facilities~\cite{akre_commissioning_2008}.

An interesting future application of emission from patterned surfaces is the possibility to obtain transverse electron beam density modulation and shaping.
Since electrons are only emitted near the groove edges, structures could be engineered to generate nanoscale beamlets and density modulations.
Using linear optics to perform an emittance exchange~\cite{nanni_nanomodulated_2018} this modulation can be transferred into time and used to drive temporal patterning in the beam which is of interest to coherent x-ray light sources~\cite{graves_intense_2012}.
Also, a round beam~\cite{derbenev_untitled_1998, brinkmann_low_2001} could be generated to average out the emittances in both directions.

To further reduce the emittance from plasmonic photocathodes, emission from a flat surface would be required, for example by using a plasmonic lens, as described in~\cite{durham_plasmonic_2019}, where surface plasmon interference is used to produce large and instantaneous field enhancements in pattern-free areas well below one micrometer.
The implementation of this design could improve the transverse brightness from metal cathodes even further over the state-of-the-art.

%TC:ignore
\section{Acknowledgements}
C.M.P. acknowledges US NSF Award PHY-1549132, the Center for Bright Beams and the US DOE SCGSR
program. D.F. acknowledges support from the Office of Science, Office of Basic Energy Sciences, of the U.S. Department of Energy, under Contract No.~DE-AC02-05CH11231.
Work at the Molecular Foundry was supported by the Office of Science, Office of Basic Energy Sciences, of the U.S. Department of Energy, under Contract No. DE-AC02-05CH11231.
D.B.D. and A.M.M. acknowledge support from NSF Science and Technology Center on Real-Time Functional Imaging (STROBE) under Grant No.~DMR-1548924, which included funds for building the laser transport line in the DC photoemission test stand used in this work. 
%TC:endignore

\appendix*

\section{Sensitivity of Reflectance to Laser Angle of Incidence}

\begin{figure}[t]
    \centering
    \includegraphics[width=\linewidth]{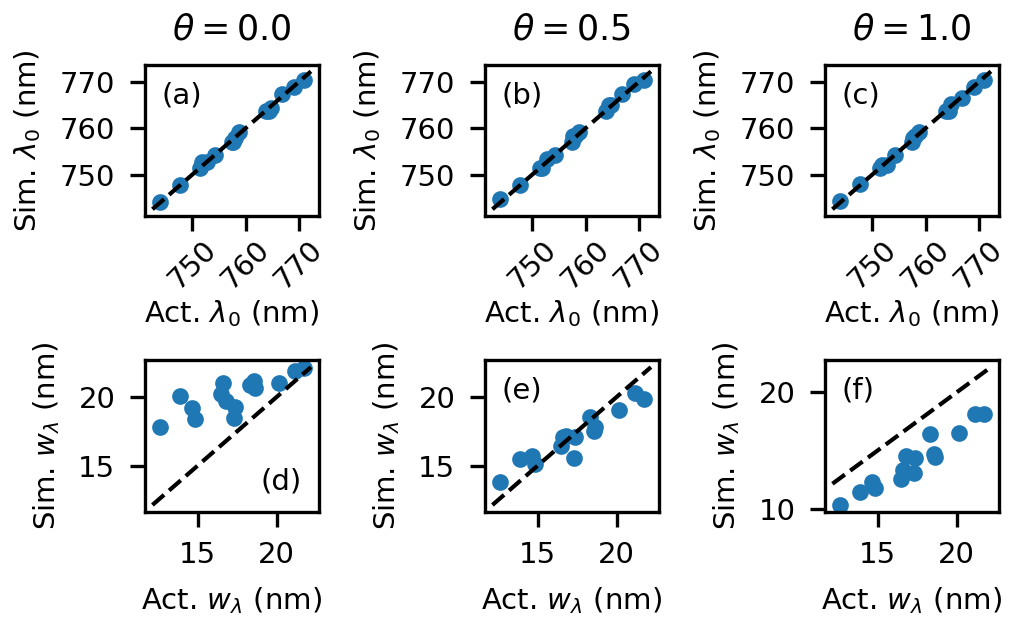}

    \caption{
    Variation of the modeled nanogroove reflectivity as laser angle of incidence is varied.
    Parameters of the reflectivity peaks compared with measurements (as in Fig. \ref{fig:reflectance-spectra}b and \ref{fig:reflectance-spectra}c) are shown for a few angles compared with the best-fit value (subplots b and e).}
    \label{fig:tilt-comparison}
\end{figure}

In Sec. \ref{sec:fabrication}, laser angle of incidence is used as one of the free parameters when fitting the FDTD simulations to measurements of reflectivity (i.e. Fig. \ref{fig:reflectance-spectra}).
Requiring the angle of incidence to be the same across all 16 nanogroove samples on the wafer, a best-fit value of $\theta=0.5$ was found and the best-fit simulations agreed closely with our measurements.
To better understand the effect of this angle on our results, we repeat the analysis, but for several fixed angles of incidence.
These results are summarized in Fig. \ref{fig:tilt-comparison}.

The main impact of changing the angle of incidence is to change the width of the absorption peaks without affecting their location.
This suggests that groove width, which does change the location of the peak and is the only remaining free parameter, can be accurately predicted from data even when there is uncertainty in the cathode's tilt.
Confirming this, the best-fit groove width varies on average by only 5\% as the angle is varied in Fig. \ref{fig:tilt-comparison}.

Additionally, we proposed in Sec. \ref{sec:nonlinear-photoemission} that aligning the absorption peak with the spectrum of the driving laser is important to optimize the performance of the cathode.
%This can be seen in the correlation between the amount of overlap between the laser's spectrum and the nanogroove's absorption in Fig. \ref{fig:nonlinear-yield}f.
Since the position of the cathode's absorption peak does not change much with the angle of incidence, the nonlinear yield should be equally robust.
This is a useful property for applications where engineering constraints place limits on how well the laser's angle of incidence can be known.

%TC:ignore
\bibliography{nanogrooves}% Produces the bibliography via BibTeX.
%TC:endignore

\end{document}